\documentclass[final,3p,times,twocolumn]{elsarticle}

\usepackage{amssymb}
\usepackage{amsmath}
\usepackage{makecell}
\usepackage{xcolor}
\usepackage{float}
\usepackage{geometry}
\journal{Surfaces and Interfaces}

\begin{document}

\begin{frontmatter}

\title{Adhesion Energy of Phosphorene on Different Pristine and Oxidized Metallic Substrates}

\author[inst1]{Matteo Vezzelli}
\author[inst2]{Carsten Gachot}
\author[inst1]{M. Clelia Righi}
\affiliation[inst1]{organization={Department of Physics and Astronomy, University of Bologna}, 
            city={Bologna},
            postcode={40127}, 
            country={Italy}}
\affiliation[inst2]{organization={Institute of Engineering Design and Product Development, TU Wien}, 
            city={Wien},
            postcode={1060}, 
            country={Austria}}

\begin{abstract}
Black phosphorus and its single-layer constituent, phosphorene, have emerged as promising two-dimensional materials with remarkable tribological properties. However, recent experimental investigations revealed that the their lubricating capabilities can change with the substrate. The present computational study employs density functional theory calculations to quantify the adhesion energy of both pristine and oxidized phosphorene monolayers on various metallic substrates (aluminum, copper, iron, and chromium) and their corresponding oxides ($\mathrm{Al_2O_3}$, $\mathrm{Cu_2O}$, $\mathrm{Fe_2O_3}$, and $\mathrm{Cr_2O_3}$), correlating these interfacial property with experimentally observed tribological performance. Results demonstrate that oxidized phosphorene presents higher adhesion to all substrates with respect to pristine phosphorene, attributed to favorable interactions between oxygen non-bonding states and substrate empty states. Adhesion is systematically more favorable on pristine metals than on their corresponding oxides, with chromium and iron showing particularly strong interactions due to partially filled 3d orbitals. This result is consistent with the coefficient of friction decrease observed in tribological experiments after scratching the iron substrate, thus removing the outermost oxide layer. Charge redistribution correlates with the adhesion and electronic structure analyses reveal system-dependent interfacial bonding characteristics, with some configurations inducing metallic character in phosphorene. These findings provide fundamental insights into substrate-dependent lubricating properties of black phosphorus, highlighting the key role of layer-substrate adhesion.
\end{abstract}


\end{frontmatter}

\section{Introduction}

Black phosphorus (BP), the most stable crystalline allotrope of phosphorus, has emerged as one of the most promising two-dimensional (2D) materials since its rediscovery as a monolayer system in 2014. This material has joined the expanding family of post-graphene 2D materials, which includes transition metal dichalcogenides, hexagonal boron nitride, and MXenes \cite{ling2015renaissance, marquis2022nanoscale}.

Phosphorene, a single layer exfoliated from bulk BP, presents unique electronic properties that have attracted considerable attention, owing to its direct bandgap of 0.3 eV \cite{kou2015phosphorene, carvalho2016phosphorene}, which can be systematically tuned by varying the number of layers \cite{low2014tunable} or applying in-plane strain \cite{rodin2014strain}. Beyond its electronic characteristics, phosphorene exhibits remarkable mechanical and tribological properties attributed to its anisotropic Young's modulus \cite{tao2015mechanical} and its exceptional ability to achieve ultralow friction when employed as a lubricant additive in aqueous solutions \cite{wang2018superlubricity}.

The superior tribological performance of phosphorene as both filler and composite material \cite{wang2018superlubricity, cui2017atomic, lv2018self, wang2018black} originates from the ability, common to other layered materials, to reduce the surface energy of the substrate \cite{losi2023modeling, restuccia2016tribochemistry,marchetto2017surface} and in the weak interlayer interactions \cite{losi2020superlubricity,philippon2007experimental, de2015tribochemistry}, which reduce interfacial adhesion and friction. However, recent experimental investigations have shown that BP performance as solid lubricant is deeply influenced by the substrate \cite{vezzelli2024different}.

From a structural perspective, BP possesses a distinctive atomic arrangement that governs its tribological properties. Each phosphorus atom contains five valence electrons in a $3s^23p^3$ configuration, forming four $sp^3$ hybridized orbitals. Three orbitals participate in bonding with adjacent phosphorus atoms, while the fourth hosts a lone electron pair, resulting in a tetrahedral geometry that produces a characteristic puckered honeycomb lattice structure. This arrangement exhibits pronounced structural anisotropy: the material displays \textit{zigzag} and \textit{armchair} configurations along the \textit{x}- and \textit{y}-directions, respectively, while exhibiting a hexagonal structure along the \textit{z}-direction. Each phosphorene layer comprises two atomic sub-layers connected by two distinct types of P-P bonds: intra-planar bonds within each sub-layer and weaker inter-planar bonds between sub-layers, as illustrated in the Supplementary Material (SM) \cite{losi2020superlubricity, benini2023interaction, zhang2021semiconducting, boddula2020black, losi2023modeling}.

\begin{figure}[H]
\centering
\includegraphics[width=\linewidth]{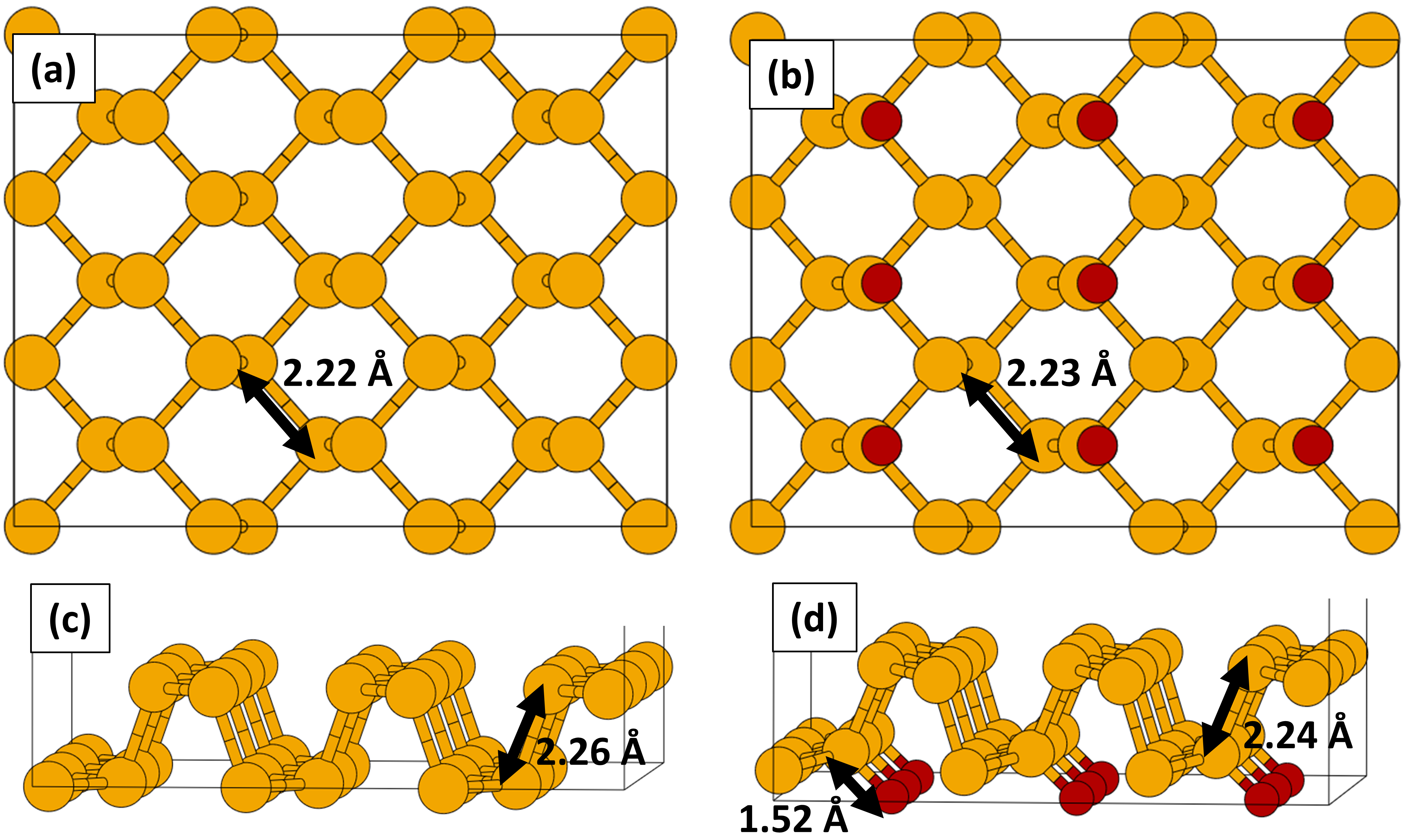}
\caption{Comparison between pristine and oxidized phosphorene structures. (a) Bottom view and (c) side view of
pristine phosphorene. (b) Bottom view and (d) side view of oxidized phosphorene with 25\% of oxygen coverage, determined as the most stable configuration for oxidized phosphorene \cite{benini2023interaction}.}\label{figure1}
\end{figure}

The weak interlayer van der Waals interactions enable the mechanical exfoliation of bulk BP into few-layer or monolayer structures, analogous to the well-established exfoliation of graphite to produce graphene. However, these reduced-dimensional BP structures suffer from intrinsic instability \cite{abellan2017fundamental,island2015environmental} and consequently exhibit high reactivity toward oxygen, moisture, and light \cite{kuntz2017control, favron2015photooxidation, luo2016surface, hu2017water, zhou2016light, huang2016interaction}. Paradoxically, the formation of phosphorus oxides resulting from BP degradation has been demonstrated to enhance friction and wear performance, as observed in atomic force microscopy (AFM) experiments \cite{wu2018black} and when BP nanoflakes are incorporated as solid fillers in polytetrafluoroethylene (PTFE) \cite{wang2022effect}. Consequently, this computational study investigates both pristine and oxidized phosphorene monolayers, as depicted in Figure \ref{figure1}.

The substrate-dependent tribological behavior of BP was systematically investigated in the study "A Different Perspective on the Solid Lubrication Performance of Black Phosphorus: Friend or Foe?" \cite{vezzelli2024different}, which demonstrated that BP's performance varies significantly with substrate material and contact conditions. That work identified adhesion between BP and the substrate as a critical factor determining its effectiveness as a solid lubricant. Building upon these experimental findings, the present computational study extends this analysis by quantifying the adhesion energy of BP on the same metal substrates (steel, aluminum, copper, and iron) through density functional theory (DFT) calculations.

The computational modeling focuses on the most stable crystallographic surfaces of the metal substrates examined in the experimental investigations. Additionally, the study considers the primary oxide phases for each metal. Pristine aluminum naturally develops a protective $\mathrm{Al_2O_3}$ (alumina) passivation layer that prevents further oxidation \cite{jayatissa2019tribological}, while on steel and iron surfaces, $\mathrm{Fe_2O_3}$ (hematite) represents the predominant oxide phase \cite{ali1969oxidation, vernon1953oxidation, davies1954oxidation, huang2016surface}. For copper, $\mathrm{Cu_2O}$ (cuprite) constitutes the primary oxide at temperatures up to 300°C, consistent with the experimental conditions employed in the referenced tribological tests \cite{vezzelli2024different, lee2016oxidation, greenwood1997chemistry}. The analysis additionally incorporates $\mathrm{Cr_2O_3}$ (chromia) to account for the thin oxide layer that naturally forms on the chromium-bearing steel (AISI 52100) utilized in the experimental studies \cite{cho1999novel, huang2023atomistic}.

This computational investigation aims to provide quantitative insights into the substrate-dependent adhesion behavior of BP, thereby elucidating the fundamental mechanisms underlying its variable tribological performance across different material systems.

\section{Computational details}

In this study, spin-polarized DFT calculations were performed following the plane-wave and pseudopotential approach as implemented in Quantum Espresso software \cite{giannozzi2009quantum}. The electronic wave functions were expanded on a plane-wave basis set, with ultrasoft pseudopotentials employed to describe the ionic species, including the core electrons \cite{vanderbilt1990soft}. The Generalized Gradient Approximation (GGA) within the Perdew-Burke-Ernzerhof (PBE) parameterizations was used as the exchange-correlation functional \cite{perdew1996generalized}. A Gaussian smearing of 0.005 Ry was employed to better describe the occupations of the electronic states around the Fermi level. Optimizations were carried out using the BFGS algorithm \cite{shanno1970conditioning} and stopped when the total energy and forces converged below the threshold of $10^{-4}$ Ry and $10^{-3}$ Ry/Bohr, respectively. The convergence criteria for the self-consistent electronic (SCF) loop were set to $10^{-6}$ Ry. For every system considered, the kinetic energy cut-off to expand the wave function and the charge densities were set to 50 Ry and 400 Ry, respectively. All systems were constructed with at least 15 Å of vacuum between vertical replicas.

The sampling of the Brillouin zone was performed without the inclusion of the $\Gamma$-point, using an appropriately converged Monkhorst-Pack grid of K-points, ensuring a total energy accuracy within 1 meV/atom compared to calculations with a denser grid. Equivalent reduced grids were employed to sample the corresponding most stable surfaces.

London dispersion forces were accounted for by adopting the D2 scheme \cite{grimme2006semiempirical}, using 0.75 for the s6 scaling factor as suggested in the original paper. This dispersion scheme has yielded very good results for computing properties of BP and phosphorene interlayer energy \cite{shulenburger2015nature}.

Since DFT faces significant difficulties in accurately modeling the electronic structure and magnetic properties of transition-metal oxides, the DFT+U method was employed to better capture the strong localization of the \textit{d} orbitals in these systems, accounting for the on-site Coulomb interactions \cite{naveas2023first}. The Hubbard parameters were determined empirically, with iterative adjustments made to optimize structural parameters and band gap values, ensuring an accurate description of the transition-metal oxides. The values used were 4.8 eV for the 3\textit{d} orbitals of $\mathrm{Fe_2O_3}$, 7.0 eV for the 3\textit{d} orbitals of $\mathrm{Cr_2O_3}$, and 11.0 eV for the 3\textit{d} and 2\textit{p} orbitals of $\mathrm{Cu_2O}$. The Hubbard plots for the bulk materials are shown as SM.

Table \ref{table1} presents a comprehensive overview of the computational details used for bulk and substrate calculations. It includes the optimized bulk lattice parameters, calculated surface energies corresponding to the most stable surfaces and terminations, and the number of substrate layers used. For pristine phosphorene, the calculated lattice parameters and surface energy are consistent with ref. \cite{dhanabalan2017emerging}. The surface energies of Al(111), Cu(111), and Fe(110) agree with previous DFT calculations \cite{jain2013commentary, kiejna2001first, vitos1998surface}. For $\mathrm{Al_2O_3}$(0001), both Al-terminated and O-terminated configurations were benchmarked against refs. \cite{chiang2016dft, ramogayana2021density, sun2006structure}. The $\mathrm{Cu_2O}$(111) O-terminated surface structure and energetics are consistent with refs. \cite{zhang2018atomistic, kokalj2015density, gao2020surface}, while $\mathrm{Fe_2O_3}$(0001) Fe-terminated and O-terminated surfaces were compared with refs. \cite{huang2016surface, nguyen2013water, rohrbach2004ab}. The Cr(110) surface energy matches literature values \cite{jain2013commentary, vitos1998surface}, as does the Cr-terminated $\mathrm{Cr_2O_3}$(0001) surface \cite{sun2006structure, rohrbach2004ab, costa2009ab}. 

\begin{table*}[h]
\centering
\footnotesize
\begin{tabular}{lcccccccc}
\hline
Material & Bulk cell & $a$ (\AA) & $b$ (\AA) & $c$ (\AA) & \makecell{Surface\\cell} & \makecell{Surface\\plane} & \makecell{Surface\\energy\\(J/m$^2$)} & \makecell{n. of\\layers} \\
\hline
Ph & orthorhom. & 3.33 & 4.33 & 10.49 & orthorhom. & [001] & 0.18 & 1 \\
Ph-ox & orthorhom. & 3.33 & 4.33 & 10.49 & orthorhom. & [001] & 0.12 & 1 \\
Al & cubic FCC & 4.02 & 4.02 & 4.02 & hexagonal & [111] & 0.75 & 4 \\
Al$_2$O$_3$ Al-term & hexagonal & 4.79 & 4.79 & 13.03 & hexagonal & [0001] & 2.11 & 6 \\
Al$_2$O$_3$ O-term & hexagonal & 4.79 & 4.79 & 13.03 & hexagonal & [0001] & 3.83 & 6 \\
Cu & cubic FCC & 3.59 & 3.59 & 3.59 & hexagonal & [111] & 1.04 & 4 \\
Cu$_2$O O-term & cubic & 4.33 & 4.33 & 4.33 & hexagonal & [111] & 1.15 & 5 \\
Fe & cubic BCC & 2.84 & 2.84 & 2.84 & primitive & [110] & 2.33 & 3 \\
Fe$_2$O$_3$ Fe-term & hexagonal & 5.11 & 5.11 & 13.76 & hexagonal & [0001] & 1.30 & 6 \\
Fe$_2$O$_3$ O-term & hexagonal & 5.11 & 5.11 & 13.76 & hexagonal & [0001] & 2.36 & 6 \\
Cr & cubic BCC & 2.82 & 2.82 & 2.82 & primitive & [110] & 3.60 & 3 \\
Cr$_2$O$_3$ Cr-term & hexagonal & 5.15 & 5.15 & 13.63 & hexagonal & [0001] & 1.57 & 6 \\
\hline
\end{tabular}
\normalsize
\caption{Optimized parameters for bulk and substrate calculations, where $a$, $b$, and $c$ refer to the bulk lattice vectors. Abbreviations: Ph, phosphorene; Ph-ox, oxidized phosphorene.}
\label{table1}
\end{table*}

\begin{table*}[h]
\centering
\footnotesize
\begin{tabular}{lccc|ccc}
\hline
& \multicolumn{3}{c|}{Phosphorene (pristine and oxidized)} & \multicolumn{3}{c}{Substrate} \\
\hline
Material & Supercell & Mismatch on $a$ & Mismatch on $b$ & Supercell & Mismatch on $a$ & Mismatch on $b$ \\
\hline
Al / Ph & 4$\times$1 & 1.9\% & 3.4\% & 5$\times$1 & 1.3\% & 3.4\% \\
Al / Ph-ox & 4$\times$1 & 2.4\% & 3.4\% & 5$\times$1 & 1.8\% & 3.4\% \\
Al$_2$O$_3$ Al-term / Ph & 3$\times$2 & 2.3\% & 4.9\% & 2$\times$1 & 1.9\% & 1.6\% \\
Al$_2$O$_3$ Al-term / Ph-ox & 3$\times$2 & 2.2\% & 4.3\% & 2$\times$1 & 2.0\% & 2.1\% \\
Al$_2$O$_3$ O-term / Ph & 3$\times$2 & 0.1\% & 4.4\% & 2$\times$1 & 4.2\% & 0.9\% \\
Al$_2$O$_3$ O-term / Ph-ox & 3$\times$2 & 0.3\% & 4.9\% & 2$\times$1 & 4.6\% & 1.5\% \\
Cu / Ph & 5$\times$1 & 3.1\% & 4.9\% & 7$\times$1 & 3.0\% & 4.9\% \\
Cu / Ph-ox & 5$\times$1 & 3.5\% & 4.9\% & 7$\times$1 & 3.3\% & 4.9\% \\
Cu$_2$O O-term / Ph & 2$\times$7 & 3.4\% & 1.1\% & 1$\times$3 & 5.0\% & 2.3\% \\
Cu$_2$O O-term / Ph-ox & 2$\times$7 & 3.0\% & 1.0\% & 1$\times$3 & 5.0\% & 2.3\% \\
Fe / Ph & 6$\times$2 & 0.2\% & 4.7\% & 5$\times$3 & 0.3\% & 0.9\% \\
Fe / Ph-ox & 6$\times$2 & 0.1\% & 2.6\% & 5$\times$3 & 0.5\% & 1.3\% \\
Fe$_2$O$_3$ Fe-term / Ph & 3$\times$2 & 2.4\% & 1.0\% & 2$\times$1 & 0.7\% & 0.3\% \\
Fe$_2$O$_3$ Fe-term / Ph-ox & 3$\times$2 & 2.6\% & 1.2\% & 2$\times$1 & 0.5\% & 0.4\% \\
Fe$_2$O$_3$ O-term / Ph & 3$\times$2 & 2.3\% & 1.0\% & 2$\times$1 & 0.7\% & 1.7\% \\
Fe$_2$O$_3$ O-term / Ph-ox & 3$\times$2 & 2.6\% & 1.2\% & 2$\times$1 & 1.0\% & 1.9\% \\
Cr / Ph & 7$\times$3 & 0.9\% & 4.4\% & 6$\times$5 & 0.2\% & 2.1\% \\
Cr / Ph-ox & 7$\times$3 & 0.4\% & 4.4\% & 6$\times$5 & 0.8\% & 2.1\% \\
Cr$_2$O$_3$ Cr-term / Ph & 3$\times$2 & 2.9\% & 1.6\% & 2$\times$1 & 0.8\% & 0.2\% \\
Cr$_2$O$_3$ Cr-term / Ph-ox & 3$\times$2 & 3.1\% & 1.8\% & 2$\times$1 & 0.7\% & 0.4\% \\
\hline
\end{tabular}
\normalsize
\caption{Summary of constructed supercells for phosphorene-substrate interactions. Abbreviations: Ph: phosphorene, Ph-ox: oxidized phosphorene.}
\label{table2}
\end{table*}

For this study, calculations focused solely on monolayer phosphorene, as the primary interest was in determining  the capability of the solid lubricant to adhere on different metal substrates. Other computational studies, where bilayers were considered, have addressed the effects on the cleavage and shear strength of the metallic interfaces \cite{losi2023modeling} and interlayer sliding properies \cite{benini2023interaction}.

The $\mathrm{Cr_2O_3}$(0001) surface was chosen for the calculations, although the (1012) surface is more stable \cite{sun2006structure}. This choice was driven by computational constraints, as constructing a supercell with the (1012) surface would have required an extremely large model size. The (0001) surface, being the second in order of stability, was found to be an acceptable approximation, additionally facilitating direct comparisons with $\mathrm{Fe_2O_3}$ surface.

For each metal oxide substrate, two possible surface terminations were considered by cleaving the structures along the \textit{z}-axis and exposing outermost atomic layers of metal atoms, "M-term", or oxygen atoms, "O-term". The relaxed geometries of these different surface terminations are shown as SM.

After obtaining the optimized structures of the bulk and substrates, orthorhombic supercells were constructed by placing pristine and oxidized phosphorene above each of the substrates at different lateral positions, with an initial average distance of 2.0 \AA. The most representative system chosen was the one with the lowest adhesion energy among the different lateral positions. The supercell was then relaxed according to the previously described criteria, maintaining a mismatch on the \textit{a} and \textit{b} vectors not exceeding 5\% relative to the optimized vectors of the isolated systems. 

\begin{figure*}[!t]
\centering
\includegraphics[width=0.8\linewidth]{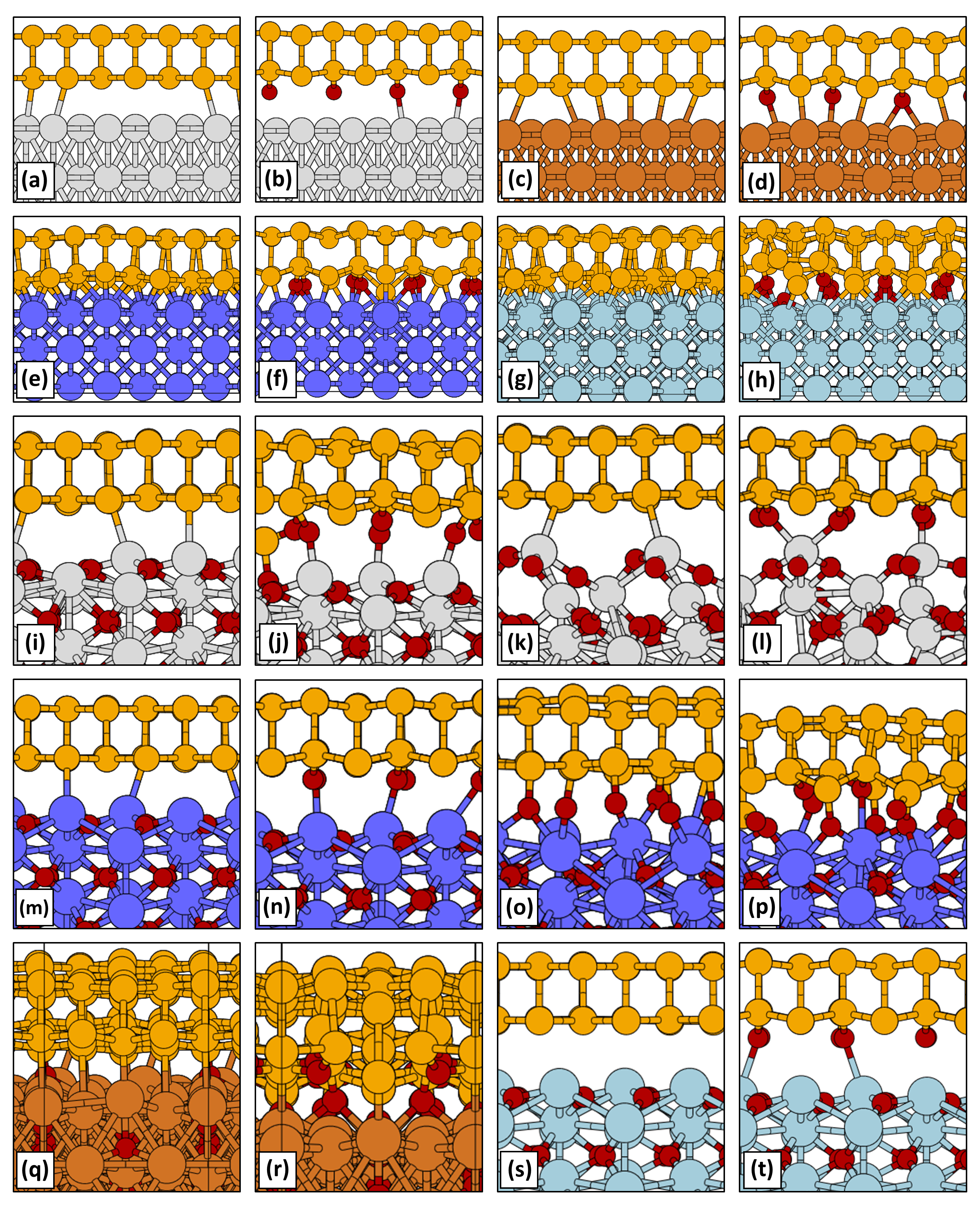}
\caption{Relaxed geometries of the supercells: (a, b) pristine and oxidized phosphorene on aluminum, (c, d) pristine and oxidized phosphorene on copper, (e, f) pristine and oxidized phosphorene on iron, (g, h) pristine and oxidized phosphorene on chromium, (i, j) pristine and oxidized phosphorene on $\mathrm{Al_2O_3}$ Al-terminated, (k, l) pristine and oxidized phosphorene on $\mathrm{Al_2O_3}$ O-terminated, (m, n) pristine and oxidized phosphorene on $\mathrm{Fe_2O_3}$ Fe-terminated, (o, p) pristine and oxidized phosphorene on $\mathrm{Fe_2O_3}$ O-terminated, (q, r) pristine and oxidized phosphorene on $\mathrm{Cu_2O}$ O-terminated, (s, t) pristine and oxidized phosphorene on $\mathrm{Cr_2O_3}$ Cr-terminated. Color code: P atoms are shown in yellow, Al in grey, Fe in purple, Cu in orange, Cr in cyan, and O in red.}\label{figure2}
\end{figure*}

Table \ref{table2} presents a summary of the constructed supercells, while the relaxed supercells for each system are shown in Figure \ref{figure2}.

The surface energy was calculated using Equation \ref{equation1}:

\begin{equation}
E_{\mathrm{surface}} = \frac{E_{\mathrm{sub}} - n \cdot E_{\mathrm{bulk}}}{2A}
\label{equation1}
\end{equation}

where $E_{\mathrm{sub}}$ and $E_{\mathrm{bulk}}$ are the total energies of the substrate and of the bulk, respectively, \textit{n} is the ratio between the number of atoms in the substrate and in the bulk, and \textit{A} is the area of the supercell. For phosphorene, the monolayer was considered for the calculation of $E_{\mathrm{sub}}$. For oxidized phosphorene, the bulk structure was constructed by replicating the most stable oxidized monolayer configuration along the out-of-plane direction. $E_{\mathrm{bulk}}$ was obtained from this multilayer oxidized structure, ensuring consistency with the oxidation pattern of the monolayer system.

The adhesion energy of pristine and oxidized phosphorene to the substrate was evaluated as:

\begin{equation}
E_{\mathrm{adh}} = \frac{E_{\mathrm{tot}} - E_{\mathrm{ph}} - E_{\mathrm{sub}}}{A}
\label{equation2}
\end{equation}

where $E_{\mathrm{tot}}$ is the energy of the whole system, while $E_{\mathrm{ph}}$ and $E_{\mathrm{sub}}$ are the energies of the isolated phosphorene and substrate, respectively, and \textit{A} is the in-plane area of the supercell.

To quantify the structural differences between the interacting structures in the supercell and their isolated counterparts, the Root Mean Square Deviation (RMSD) was employed through a custom Python script. RMSD, defined as $ \sqrt{ \frac{ \sum_{i} d_{i}^{2} }{ n } }$ , where $d_i$ represents the distance between each of the \textit{n} pairs of equivalent atoms in two optimally superimposed structures, is commonly used in protein structure analysis and MD simulations for expressing structural similarity \cite{debefve2021systematic, sharma2014accuracy}. 

 An RMSD value of 0 \AA\ indicates identical structures, with increasing values reflecting greater structural differences \cite{carugo2001normalized}. RMSD values are considered reliable indicators of variability when applied to highly similar structures, providing a quantitative measure of geometrical resemblance \cite{carugo2001normalized}. In this study, visual inspection of the phosphorene-substrate interactions revealed noticeable but not excessive distortions. Therefore, RMSD evaluation was considered an appropriate approximation for quantifying the degree of structural distortion.

To analyze the electronic charge redistribution resulting from the interaction between phosphorene and the substrate, the charge displacement was calculated by subtracting the charge densities of the isolated constituents from that of the interface. The resulting $\Delta\rho(x,y,z)$ was then integrated along the \textit{xy}-plane to obtain the planar average charge density redistribution, $\Delta\rho(z)$. Finally, the total amount of electronic charge redistributed per surface unit ($Q_{red}$) was computed by integrating the absolute value of $\Delta\rho(z)$ along the \textit{z}-axis and divided by the supercell area \cite{wolloch2018interfacial}.

To investigate the electronic structure of phosphorene, Projected Density of States (PDOS) calculations were performed using a denser \textit{K}-points grid. The PDOS provide insights into the contributions of specific orbitals to the electronic states at each energy level, allowing for a detailed examination of the energy occupation around the Fermi level. In this study, the PDOS of the \textit{p} orbitals in pristine and oxidized phosphorene were calculated and compared with that of the isolated structure.

\section{Results and discussion}

In order to determine the optimal structure of oxidized phosphorene, a series of calculations were performed. This involved placing an oxygen atom 1.5 \AA\ away from each phosphorus atom and testing various configurations. Combinations of one- and two-sided oxygen adsorption were explored at different coverages (100\%, 75\%, 50\%, and 25\%). As reported in the literature \cite{benini2023interaction}, the most favorable configuration, exhibiting the lowest oxygen adsorption energy, corresponded to 25\% coverage. A comparison between pristine and oxidized phosphorene is shown in Figure \ref{figure1}. It can be seen that the oxidized geometry exhibits greater distortion compared to the pristine structure, as evidenced by differences in bond lengths. Notably, as indicated in Table \ref{table1}, oxidized phosphorene exhibits a lower surface energy compared to its pristine counterpart (0.12 $J/m^2$ and 0.18 $J/m^2$, respectively). The higher stability of the oxidized form \cite{tran2016surface} aligns with experimental observations, which show that BP tends to oxidize spontaneously under ambient conditions \cite{zhong2023exploring}.

\begin{figure}[H]
\centering
\includegraphics[width=\linewidth]{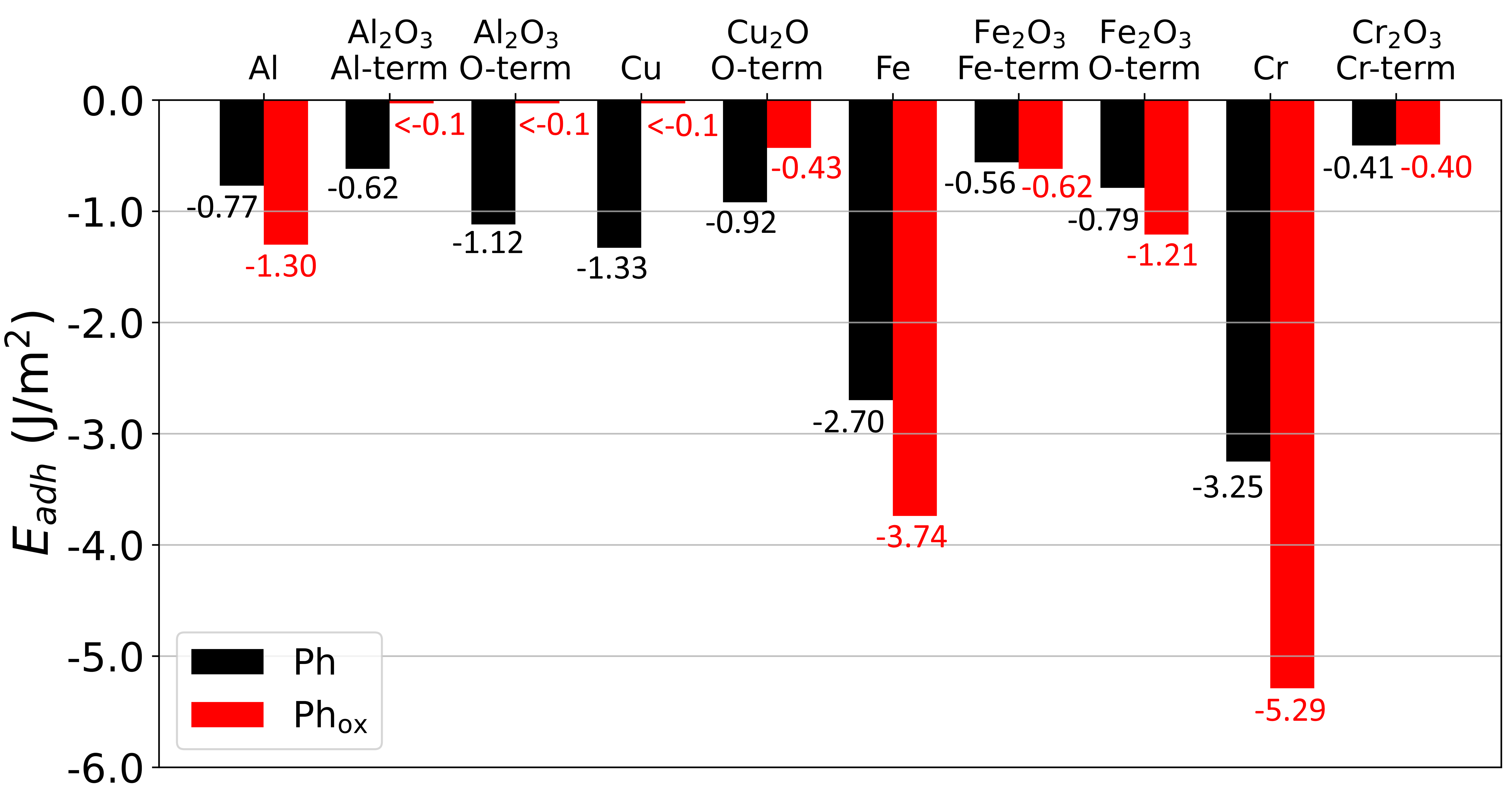}
\caption{Adhesion energies for every phosphorene-substrate system.}\label{figure3}
\end{figure}

The adhesion energies of pristine and oxidized phosphorene with various substrates were calculated to assess the strength of interaction, with the results presented in Figure \ref{figure3}. Overall, oxidized phosphorene exhibits slightly more favorable adhesion (more negative values) across all substrates. This could be attributed to the non-bonding states of the oxygen atoms, which can better interact with the empty states of the substrates. The adhesion to metallic surfaces of chromium and iron is particularly strong \cite{losi2023modeling}, especially for oxidized phosphorene. This enhanced interaction may be attributed to the partially filled 3\textit{d} orbitals of Cr and Fe atoms, which are available to form bonds with both P and O atoms. In contrast, aluminum and copper show weaker adhesion, though still favorable. However, the interaction between copper and oxidized phosphorene is nearly negligible. These differences could be due to the different electronic configurations: Al has only one electron in the 3\textit{p} orbital, while Cu has fully filled 3\textit{d} orbitals, leading to different interactions with P and O atoms.

For oxide substrates, the adhesion energies are generally less favorable (less negative values) compared to their corresponding pristine metals. This can be attributed to the reduced availability of electronic charge for bonding, as oxides typically have fewer delocalized electrons than metals. Among the oxides, $\mathrm{Fe_2O_3}$ shows the strongest adhesion, particularly with oxidized phosphorene and oxygen termination. $\mathrm{Cu_2O}$ exhibits better adhesion with pristine phosphorene, while $\mathrm{Cr_2O_3}$ shows similar values for both forms. Notably, while $\mathrm{Al_2O_3}$ presents comparable adhesion to other oxides with pristine phosphorene, its interaction with oxidized phosphorene is extremely weak.

The relationship between the calculated adhesion energies and the experimental tribological performance reported in ref. \cite{vezzelli2024different} reveals important insights. In the experiments, the best results in terms of coefficient of friction (CoF) reduction were observed for scratched iron, where the superficial oxide layer was most likely removed, while non-scratched iron exhibited poor performance. This suggests that phosphorene-substrate adhesion plays a crucial role in enhancing the lubricating properties of BP, being significantly higher on pristine iron than on oxidized iron (Figure \ref{figure3}). This observation is consistent with findings for other layered materials used as solid lubricants, such as MXenes \cite{marquis2025mxene}, and can be explained by considering that higher adhesion of the layers to the substrate prevents their peeling off during the rubbing process. Indeed, BP was found to reduce the CoF of non-scratched aluminum substrate only at the beginning of the tribological test, while after rubbing it reached the value measured for the unlubricated sample \cite{vezzelli2024different}, in agreement with the low phosphorene-$\mathrm{Al_2O_3}$ adhesion calculated in this work (Figure \ref{figure3}). The low phosphorene-$\mathrm{Cu_2O}$ adhesion may suggest an irrelevant effect of BP in reducing the CoF of copper, as observed for aluminum. However, the experimental results indicate an even worse effect of BP on copper, which was interpreted as third-body wear effect due to the poor adhesion.

\begin{figure}
\centering
\includegraphics[width=\linewidth]{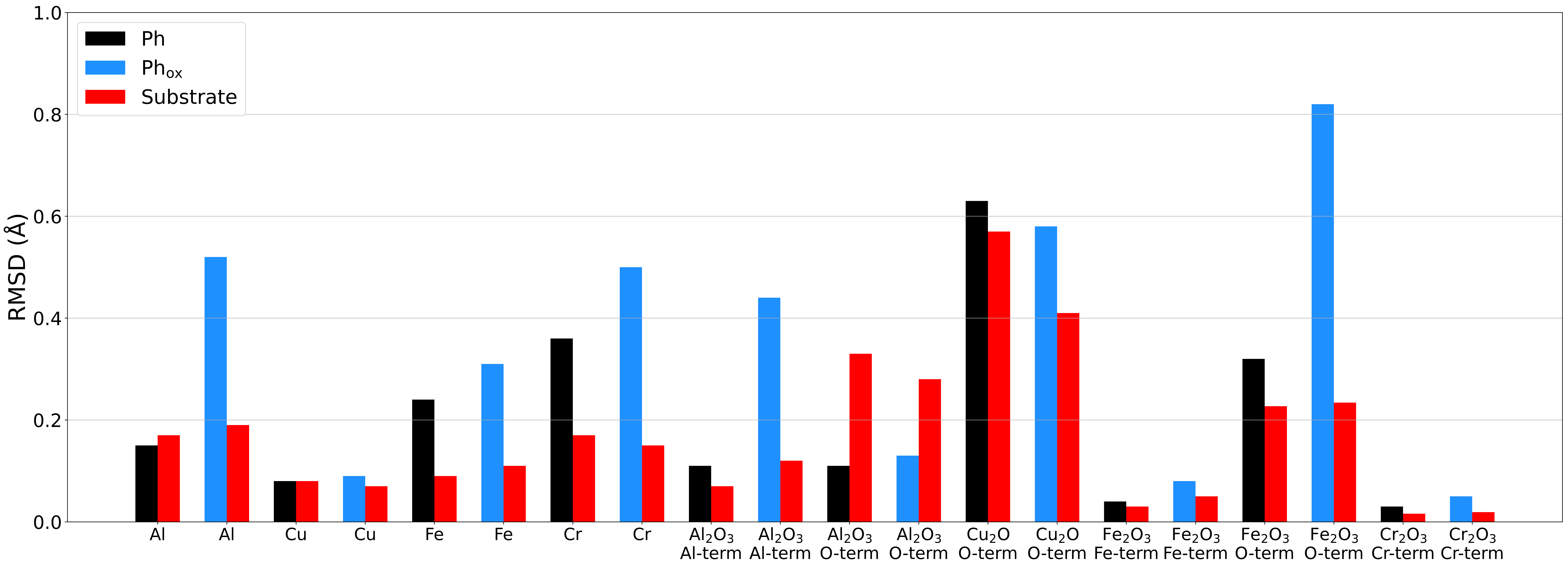}
\caption{RMSD analysis for every phosphorene-substrate system.}\label{figure4}
\end{figure}

Figure \ref{figure4} presents the evaluation of RMSD for phosphorene-substrate interactions. Analysis of the data reveals a weak correlation between RMSD values and adhesion energies. In general, oxidized phosphorene exhibits higher RMSD values compared to its pristine counterpart, aligning with its more favorable adhesion energies. This suggests that oxidized phosphorene undergoes greater structural distortion due to the more favorable interaction. The structural changes are predominantly confined to the phosphorene layer, with minimal deformation occurring in the substrate materials. Notably, $\mathrm{Cu_2O}$ stands as an exception to this trend.

\begin{figure*}
\centering
\includegraphics[width=\linewidth]{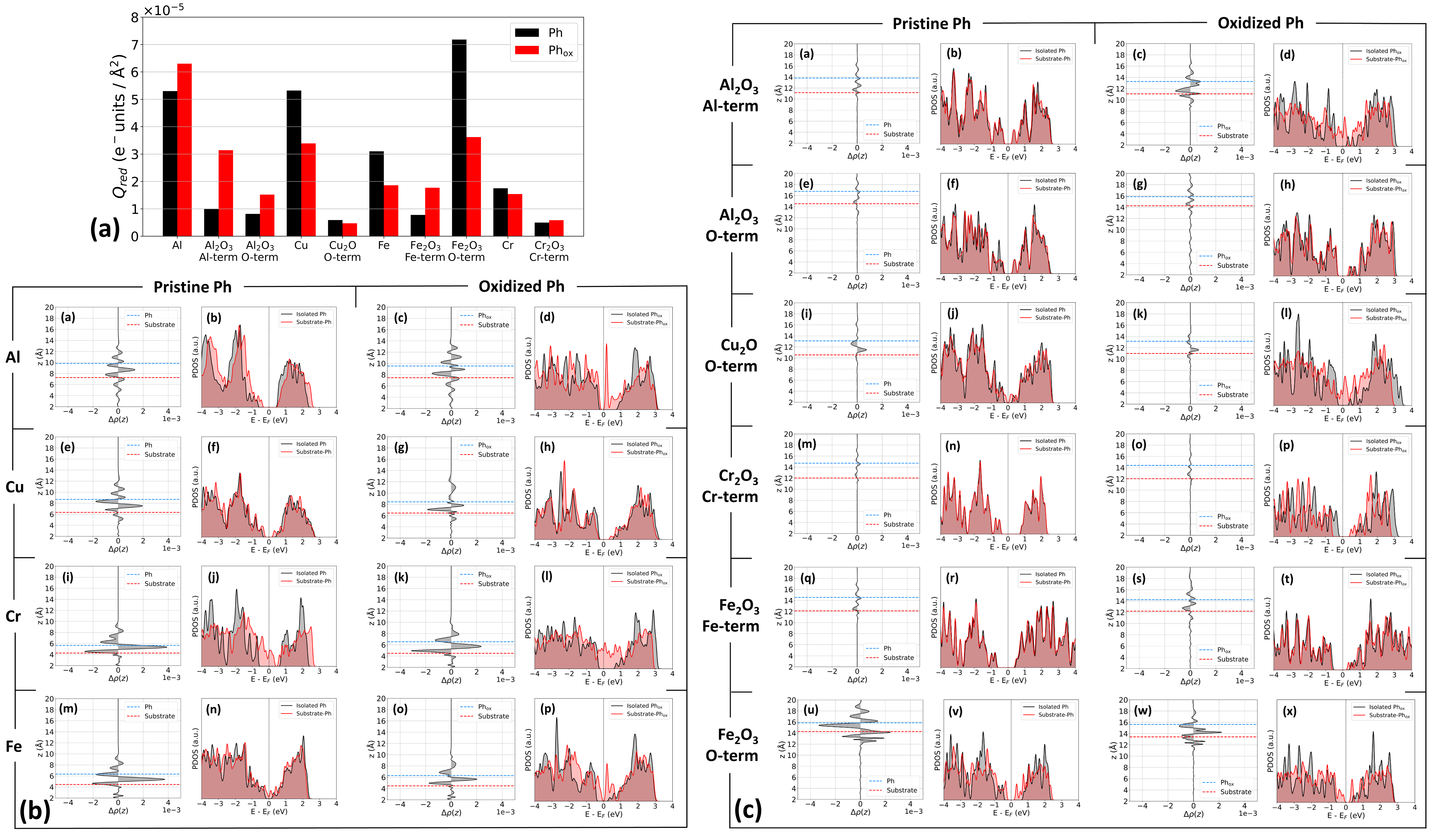}
\caption{(a) Total electron charge redistribution per surface unit ($Q_{red}$) for every phosphorene-substrate system; (b) $\Delta\rho(z)$ and PDOS analysis for phosphorene deposited on pristine metals and (c) the corresponding oxides.}\label{figure5}
\end{figure*}

The analysis of the electronic properties of the considered interfaces is reported in Figure \ref{figure5}. Panel a) presents the overall assessment of charge redistribution ($Q_{red}$) when phosphorene is deposited on the substrate. In agreement with previous studies, e.g., \cite{losi2023modeling, wolloch2018interfacial}, higher interfacial adhesion (i.e., more negative adhesion energies) is associated with higher $Q_{red}$ values. Moreover, $Q_{red}$ follows the same trend as adhesion, increasing from Al to Cu, Fe, and Cr in the case of pristine metals, while it remains very low for oxides, except for O-terminated $\mathrm{Fe_2O_3}$, where the adhesion is higher than for the other oxides. The profiles of the charge displacement, $\Delta\rho(z)$, reveal that a greater amount of electronic charge is accumulated at the interface when phosphorene is deposited on pristine metals (Figure \ref{figure5}b) than oxides (Figure \ref{figure5}c), attributed to the stronger chemisorption on the former substrates.
From the PDOS analysis, an increase in phosphorene \textit{p}-states within the band gap is observed in some cases after interaction with the substrate. In these cases, phosphorene exhibits metallic character \cite{pan2016monolayer}. In some instances, metallic behavior emerges only for oxidized phosphorene, e.g., Figure \ref{figure5}b inset (d), whereas in other cases it is observed for both pristine and oxidized phosphorene, e.g., insets (j,l). Conversely, in many other cases the insulating character is preserved. Moreover, no strong correlation is found between the emergence of metallic behavior in phosphorene and the $Q_{red}$ values, as evidenced by the comparison between Figure \ref{figure5}b insets (j,l) and the $Q_{red}$ values reported for Cr in Figure \ref{figure5}a.

\section{Conclusion}

DFT calculations were performed to determine the adhesion energies of both pristine and oxidized phosphorene on different metal substrates, also considering the corresponding metal oxides. The results demonstrate that oxidized phosphorene generally exhibits higher adhesion to all substrates compared to its pristine counterpart. This enhanced adhesion can be attributed to more favorable interactions between the non-bonding states of oxygen atoms and the empty electronic states of the substrates. 

The RMSD structural analysis revealed that oxidized phosphorene exhibits higher structural distortion, consistently with its more stronger adhesion to substrates. Phosphorene adhesion is always higher on pristine metal surfaces than on metal oxides. This difference stems from the reduced availability of electronic charge for bonding in metal oxides, due to increased localization within the oxide structure. 

Comparison of the calculated adhesion energies with the lubricating performance of BP measured in the experiments described in ref. \cite{vezzelli2024different} reveals that stronger adhesion  correlates with reduced CoFs. Indeed the lowest CoF was measured for the iron substrate after the removal of the oxide layer by initial scratch of the sample, in line with the high adhesion calculated for pristine iron. Consistently, the nearly negligible effect of BP measured for aluminum, copper, iron, and steel samples agrees with the low adhesion predicted for phosphorene on oxidized metal surfaces. 

The charge redistribution analysis ($Q_{red}$) demonstrated that greater charge redistribution generally occurs with pristine metals compared to their respective oxides, consistent with the observed adhesion energy trends. The PDOS analysis revealed that strong interaction with the substrate, such as with pristine iron and chromium surfaces, leads to the filling of phosphorene \textit{p}-states within the band gap, transforming its character from semiconducting to metallic. 

The present computational work provides fundamental insights into the substrate-dependent lubricating performance of BP by establishing quantitative relationships between interfacial adhesion energies and experimentally observed friction behavior, demonstrating that layer-substrate adhesion is a key factor governing the effectiveness of BP as a solid lubricant.

\section*{CRediT authorship contribution statement}
\textbf{Matteo Vezzelli}: Writing - Review \& Editing, Writing - original draft, Visualization, Validation, Investigation, Formal analysis, Conceptualization. \textbf{Carsten Gachot}: Review \& Editing, Conceptualization. \textbf{Maria Clelia Righi}: Writing - Review \& Editing, Conceptualization, Supervision, Funding acquisition.

\section*{Declaration of competing interest}
The authors declare that they have no known competing financial interests or personal relationships that could have appeared to influence the work reported in this paper.

\section*{Data availability}
Data will be made available on request.

\section*{Acknowledgements}
These results are part of the project “Advancing Solid Interface and Lubricants by First Principles Material Design (SLIDE)” that has received funding from the European Research Council (ERC) under the Horizon 2020 Research and Innovation program of the European Union (Grant agreement No. 865633). Furthermore, we acknowledge the CINECA award under the ISCRA initiative, for the availability of high-performance computing resources and support.

\bibliographystyle{elsarticle-num}
\bibliography{cas-refs}

\end{document}